\documentclass[12pt]{amsart}
\usepackage{amsmath, amssymb, latexsym, amsthm, mathrsfs,algo, geometry} 

\newcommand{\nc}{\newcommand}
\nc{\len}{\operatorname{len}}
\nc{\ed}{\end{document}}
\nc{\la}{\langle}
\nc{\ra}{\rangle}
\nc{\set}[2]{\{\, #1 : #2\,\}}
\nc{\N}{\mathbb{N}}
\nc{\SGDLP}{Semigroup DLP}
\newtheorem{thm}{Theorem}[section]
\nc{\bthm}{\begin{thm}} \nc{\ethm}{\end{thm}}
\newtheorem{prop}[thm]{Proposition}
\nc{\bprp}{\begin{prop}} \nc{\eprp}{\end{prop}}
\newtheorem{fact}[thm]{Fact}
\nc{\bfct}{\begin{fact}} \nc{\efct}{\end{fact}}
\newtheorem{prob}[thm]{Problem}
\nc{\bprb}{\begin{prob}} \nc{\eprb}{\end{prob}}
\newtheorem{lem}[thm]{Lemma}
\nc{\blem}{\begin{lem}} \nc{\elem}{\end{lem}}
\newtheorem{claim}[thm]{Claim}
\nc{\bclm}{\begin{claim}} \nc{\eclm}{\end{claim}}
\newtheorem{cor}[thm]{Corollary}
\nc{\bcor}{\begin{cor}} \nc{\ecor}{\end{cor}}
\newtheorem{red}[thm]{Reduction}
\nc{\bred}{\begin{red}} \nc{\ered}{\end{red}}

\newtheorem{conj}[thm]{Conjecture}
\nc{\bcnj}{\begin{conj}} \nc{\ecnj}{\end{conj}}
\theoremstyle{definition}
\newtheorem{defn}[thm]{Definition}
\nc{\bdfn}{\begin{defn}} \nc{\edfn}{\end{defn}}
\theoremstyle{remark}
\newtheorem{rem}[thm]{Remark}
\nc{\brem}{\begin{rem}} \nc{\erem}{\end{rem}}
\newtheorem{cnv}[thm]{Convention}
\nc{\bcnv}{\begin{cnv}} \nc{\ecnv}{\end{cnv}}
\newtheorem{exam}[thm]{Example}
\nc{\bexm}{\begin{exam}} \nc{\eexm}{\end{exam}}
\nc{\bpf}{\begin{proof}} \nc{\epf}{\end{proof}}
\nc{\be}{\begin{enumerate}}
\nc{\ee}{\end{enumerate}}
\nc{\bi}{\begin{itemize}}
\nc{\itm}{\item}
\nc{\ei}{\end{itemize}}

\title[A reduction of semigroup DLP to classic DLP]{\textsf{NOTE}\\[0.5cm]A reduction of semigroup DLP to classic DLP}

\author{Matan Banin}
\email[Matan Banin]{baninmmm@gmail.com}

\author{Boaz Tsaban}
\email[Boaz Tsaban]{tsaban@math.biu.ac.il}
\urladdr[Boaz Tsaban]{http://www.cs.biu.ac.il/\~{}tsaban}

\address{Department of Mathematics, Bar-Ilan University, Ramat Gan 5290002, Israel}



\begin{document}

\begin{abstract}
We present a polynomial-time reduction of the discrete logarithm problem in any periodic (a.k.a.\ torsion)
semigroup (\SGDLP{}) to the classic DLP in a sub\emph{group} of the same semigroup.
It follows that \SGDLP{} can be solved in polynomial time by quantum computers, and that
\SGDLP{} has subexponential algorithms whenever the classic DLP in the corresponding groups has subexponential algorithms.
\end{abstract}

\mbox{}\vspace{-1.2cm}

\maketitle

\section{introduction}

For Discrete Logarithm Problem (DLP) based cryptography, it is desirable to find efficiently implementable
groups for which sub-exponential algorithms for the DLP are not
available. Thus far, the only candidates for such groups seem to be (carefully chosen) groups
of points on elliptic curves \cite{kob,victor}. Groups of invertible matrices over a finite field, proposed
in \cite{var}, where proved by Menezes and Wu \cite{menezes} inadequate for this purpose.
In their paper \cite{KKS13},  Kahrobaei, Koupparis and Shpilrain propose to use \emph{semigroups}
and, in particular, the semigroup of all matrices over a certain finite group-ring as a platform for the
Diffie--Hellman protocol.

Let $S$ be a semigroup. The \emph{order} of an element $g\in S$ is the cardinality of the set $\set{g^k}{k\in\N}$.
A semigroup $S$ is \emph{periodic} (a.k.a.\ \emph{torsion}) if each element of $S$ has finite order.
The DLP in a periodic semigroup $S$ is the problem of finding, given an element $g\in S$ and a power $h$ of $g$,
a natural number $k$ such that $g^k=h$.
\emph{\SGDLP{}} is the general problem of solving the DLP in periodic semigroups.\footnote{The nonperiodic
case is discussed briefly in Section \ref{sec:nonper} below.}

We will demonstrate that the \SGDLP{} is not harder than the classic DLP in groups. Moreover,
the DLP in a semigroup $S$ reduces, in polynomial time, to the DLP in a subgroup $G$ of $S$.
Thus, if there is a subexponential time algorithm for the DLP in subgroups of $S$ (which is the case, for example,
when $S$ is a semigroup of matrices over a finite field, by the Menezes--Wu result \cite{menezes}),
then there is one for the DLP in $S$. In particular, as the DLP in groups is efficiently solvable by quantum computers,
it follows that the \SGDLP{} is efficiently solvable by quantum computers.

\subsection*{Related work}
The \SGDLP{} is quite old. It already appears in McCurley's 1990 survey \cite{McCurley}.
In \cite{MUsh}, Myasnikov and Ushakov reduce the DLP in the semigroup proposed in \cite{KKS13} to
the DLP in a group of invertible matrices over a finite field, and deduce that the DLP in that semigroup
can be solved by quantum computers. They achieve this goal by embedding the semigroup in the semigroup of
all matrices over a finite field, and then applying Jordan Canonical Form theory to reduce the problem
to the DLP in the group of invertible matrices over the same field. Our solution shows, in particular,
that specialized methods are not necessary to solve this problem.
We have reported our solution, without details, to Myasnikov and Ushakov,
and suggested to ask experts whether this was known.
Following that, Myasnikov and Ushakov consulted Steinwandt,
who mentioned the problem to some, including Childs and Ivanyos.
Unaware of our solution, Childs and Ivanyos
came up with an independent proof that the \SGDLP{} can be solved by quantum computers \cite[Sections 1--3]{ChIv}.
Their method is different, and uses quantum computers directly.
Our solution is slightly more general, since our reduction uses only classic computational assumptions.
This is useful when quantum computers are not available, but subexponential algorithms are available
for the DLP in the relevant groups, e.g., in matrix semigroups over finite fields.

\section{A solution of the \SGDLP{} using a classic DLP oracle}

Let $G$ be a finite group. By \emph{DLP oracle for $G$} we mean an oracle that,
whenever provided with a generator $g$ of $G$ and a power $h$ of $g$, returns a natural
number $k$, of size polynomial in the relevant parameters (including the length of $g$ and the order of $G$),
such that $h=g^k$.
Note that the oracle is not provided with $G$ directly, but via its generator and the definition of multiplication in the group.

We assume that each element in the considered semigroup has a unique representation, or, equivalently for our purposes,
a canonical representative that can be computed in polynomial time.
We also assume that multiplication in the semigroup can be carried out in polynomial time.

\bigskip

The following lemma should be well known. For completeness, we include a proof.

\blem\label{lem1}
Let $S$ be a periodic semigroup, and $g$ be a member of $S$.
Let $l,n$ be minimal\footnote{It does not matter whether we first minimize $l$ and then $n$, or vice versa.}
with $g^{l+n}=g^l$,
and let $t$ be minimal with $tn>l$.
Then the set
$$G:=g^l\cdot\set{g^k}{0\leq k< n}=\set{g^{l+k}}{0\leq k< n}$$
is a cyclic group of order $n$, with neutral element $g^{tn}$
and generator $g^{tn+1}$.
Moreover, $g^{tn}=g^{sn}$ for all $s\ge t$.
\elem
\bpf
As $g^{l+n}=g^l$, the set $G$ is closed under products.
The element $g^{tn}$ is neutral: As
$$g^{tn}g^l=g^{tn+l}=g^{l+tn}=g^l,$$
we have that $g^{tn}g^lg^k=g^lg^k$
for each element $g^lg^k\in G$.
Let $s\ge t$. Then $g^{sn}=g^{tn+(s-t)n}=g^{tn}$.

Inversion: Given $g^{l+k}\in G$, let $d$ be such that $l+k+d \cong tn-l \pmod n$.
Then $$g^{l+k}g^{l+d}=g^lg^{l+k+d}=g^lg^{tn-l} = g^{tn}.$$

Generator: As $g^{tn}$ is neutral, for each element $g^a$ in $G$ we have that
$$g^a = (g^{tn})^ag^a = g^{tna+a}=(g^{tn+1})^a.\qedhere$$
\epf

Let $S$ be a semigroup and let $g$ be an element of $S$.
Let $l$, $n$, and $t$ be the numbers defined in Lemma \ref{lem1}.

\bred\label{red:gporder}
Finding $n$, using a DLP oracle for $G$.
\ered
\bpf[Procedure]
Fix a number $N$ with $N\gg l+n$.
This can be done by beginning with a fixed number $N$, and doubling it until the following procedure works.
Choose random $k\in \{\lceil N/2\rceil,\dots,N\}$, and compute $h:=g^k$.
The distribution of $h$ is statistically indistinguishable from the uniform distribution on $G$.
As $\varphi(n)/n$ is greater than $1/(e^\gamma\log\log n+3/\log\log n)$
for $n>2$ \cite[Theorem 8.8.7]{BaSh} (and is at least $1/2$ for $n=1$ or $2$),
we assume, for a while, that $h$ is a generator of $G$.

It is known that the order of a group $G$ can be computed given a generator $h$ and a DLP oracle for that group.
Briefly, this can be done, using our notation, as follows.
Choose a random $k\in \{1,\dots,N\}$, and compute $k':=\log_h(h^{k})$.
It may be that calling the oracle twice with the same input, we obtain different values of $k'$.
However, the distribution of $k'$ depends only on $k \bmod n$ and not on $k$ itself.
Thus, taking the greatest common divisor of $O(1)$ differences $k-k'$ of this kind, we will obtain $n$.

Now, in any case, $h$ generates a subgroup of $G$, whose order is found by the above-mentioned
algorithm. This order divides the order of $G$. Repeating this procedure for $O(\log\log n)$ elements $h$,
the maximum (or least common multiple) of the obtained orders will be the order of $G$.
\epf

We now find $t$, using our knowledge of $n$.
The neutral element of a group is its unique element $e$ satisfying $e=e^2$, an idempotent element.
By Lemma \ref{lem1}, we need to find the minimal $t$ such that $g^{tn}$ is an idempotent. Given that
$g^{sn}=g^{tn}$ for all $s\ge t$ and $g^{sn}\ne g^{tn}$ for $s<t$, this can be done by binary search,
as in the following algorithm.

\begin{algorithm} {Finding the minimal $t$ such that $g^{tn}$ is the neutral element of $G$}{}
$b\qlet 1$\\
\qwhile $g^{nb}\neq (g^{nb})^2$\\
$b\qlet2b$
\qend\\
$e\qlet g^{nb}$ (this is the neutral element of $G$)\\
$a\qlet \frac{b}{2}$\\
\qrepeat\\
$c\qlet \frac{a+b}{2}$\\
\qif $g^{nc}\neq e$\\ \qthen\\
$a\qlet c$\\
\qelse\\
$b\qlet c$\qfi
\quntil $b-a=1$\\
\qreturn $b$
\end{algorithm}

\brem
One can use a variation of the above algorithm, that precomputes a logarithmic number of powers $g^{2^i}$,
and replaces each power computation by one multiplication. This applies to all algorithms in this paper.
\erem


Let $g^x$ be given. There are two cases to consider.
First, assume that $g^x\in G$.
Find $t$ and $n$ as above. Compute $tn$.
Let $r=g^{tn+1}$, a generator of $G$.
Using the oracle, we obtain a number $x'$ such that $r^{x'}=g^x$.
Then $g^x=r^{x'}=g^{x'(tn+1)}$.
Take $k=x'(tn+1)$. Then $g^k=g^x$, and we are done.

The following immediate fact shows that membership in $G$ can be tested efficiently.

\blem
For each $x\in\N$, we have that $g^x\in G$ if and only if $g^ng^x=g^x$. \qed
\elem

If $g^x\notin G$, we use binary search to find the minimal $b$ such that $g^{bn}g^x\in G$, as follows.

\begin{algorithm} {Finding the minimal $b$ such that $g^{bn}g^x\in G$}{}
$b\qlet 1$\\
\qwhile $g^{bn}g^{x}\notin G$\\
$b\qlet 2b$
\qend\\
$a\qlet \frac{b}{2}$\\
\qrepeat\\
$c\qlet \frac{a+b}{2}$\\
\qif $g^{cn}g^{x}\notin G$\\ \qthen\\
$a\qlet c$\\
\qelse\\
$b\qlet c$\qfi
\quntil $b-a=1$\\
\qreturn $b$
\end{algorithm}

Similarly, if $g^k\in G$ and $k$ is known,
we can use binary search to find the maximal $c$ such that $k-cn>0$ and $g^{k-cn}\in G$.

\bred
Computing a discrete logarithm of $g^x$, using a DLP oracle for $G$.
\ered
\bpf[Procedure]
It remains to consider the case where $g^x\notin G$.
Let $r=g^{tn+1}$ be the generator of $G$.
Use the above algorithm to find  the minimal $b$ such that $g^{bn}g^x\in G$.
As $n$ is the order of $G$, for each $a$ with $g^{bn+x}=g^a$, we have that $bn+x\le a$.

Using the oracle, compute $x':=\log_r g^{bn+x}$.
Then $g^{bn+x}=g^{x'(tn+1)}$, and thus $bn+x\le x'(tn+1)$.
Note that the number $x'(tn+1)$ is known.
Using binary search, find the maximal $c$ such that
$g^{x'(tn+1)-cn}\in G$. Then $bn+x=x'(tn+1)-cn$, and thus
$x=x'(tn+1)-cn-bn$ is found.
\epf

\section{A comment on nonperiodic semigroups}\label{sec:nonper}

Our assumption on the given semigroups are natural, but
it may be still interesting to consider the case where the element $g$ of $S$ has \emph{infinite} order.
Indeed, this case was proposed, by Vladimir Shpilrain, at the conference
\emph{Algebraic Methods in Cryptography}, Ruhr Universit\"at Bochum, Germany, 2005.
In this case, the semigroup $\la g\ra$ generated by $g$ is isomorphic to the additive semigroup
$\N$, but the isomorphism may be infeasible to compute. However, as the second named author
commented in that conference, there is likely to be a strong correlation between the bitlength of $g^k$
and the power $k$ (for each fixed coding of the semigroup elements).
The following algorithm should be able to recover $k$ from $g^k$ in many cases of interest.
In this algorithm, the function $\len(g)$ may be the bitlength of the presentation of $g$ as a bitstring, or
any other reasonable length function that tends to get larger for larger powers of $g$.
Line 6 of the following algorithm should be implemented by binary search.

\begin{algorithm} {Find $k$ given $h:=g^k$, for $g$ of infinite order.}{}
Choose large $P$ and $m$, polynomial in the relevant parameters.\\
\qfor $i$ from $1$ to $m$\\
Choose a random element $r\in \{\lceil P/2\rceil,\dots,P\}$.\\
$\tilde g\qlet g^r$\\
$\tilde h\qlet h^r$\\
$k\qlet \max\set{k}{\len(\tilde g^k)\le\len(\tilde h)}$\\
\qif $g^{k}=h$, \\
\qreturn $k$
\qend
\end{algorithm}

The rational of this proposal is that a small number of bits can only code a limited number of semigroup
elements, and thus a limited number of powers of $g$.
Thus, on average, the higher the power of $g$ is, the more bits are needed to code this power.

We have tested this algorithm in the case where $G$ is the braid group, $m$ (the number of tries) is $1$ (!),
$P$ is $16$, and $\len(g)$ is the number of generators in the canonical form of $g$.\footnote{There are much better
length functions for this group. We wanted to make life hard for our algorithm.}
For several parameter settings tested, the algorithm never failed.
The algorithm did fail, occasionally, when we took $P$ to be very small, so the contribution of the random power
seems important.

It may be possible to fool this algorithm if the coding is chosen in a malicious way. The question whether
there is, for each prescribed (black-box) infinite cyclic semigroup, an efficient solution to the the DLP remains,
at present, open.

‏
\end{document}